\documentclass[aps,prd,reprint,superscriptaddress,nofootinbib,amsmath,amssymb]{revtex4-2}

\usepackage{graphicx}
\usepackage{dcolumn}
\usepackage{bm}
\usepackage{hyperref}
\usepackage{xcolor}
\usepackage{booktabs}
\usepackage{multirow}
\usepackage{slashed}

\newcommand{\etapm}{\tilde{\eta}^{\pm}}
\newcommand{\etamp}{\tilde{\eta}^{+}\tilde{\eta}^{-}}

\newcommand{\dt}{\Delta t}
\newcommand{\Ne}{N_e}
\newcommand{\Nmu}{N_\mu}
\newcommand{\NH}{\mathrm{NH}}
\newcommand{\IH}{\mathrm{IH}}
\newcommand{\Rdisc}{\mathcal{R}}
\newcommand{\NuFit}{\textsc{NuFit}~6.1}

\begin{document}

\title{Flavor Tomography of Long-Lived Neutrino-Mass Mediators:\\
Probing the Neutrino Mass Ordering at the HL-LHC}

\author{Renjie Wang}
\email{rjwang@ihep.ac.cn}
\thanks{Supported by the Internal Research Fund of the Institute of
High Energy Physics, Chinese Academy of Sciences.}
\affiliation{Institute of High Energy Physics,
  Chinese Academy of Sciences, Beijing 100049, China}

\date{\today}

%=============================================================================
\begin{abstract}
The neutrino mass ordering remains unresolved because long-baseline
oscillation measurements are entangled with the unknown CP phase
$\delta_{CP}$; a collider probe free of this degeneracy would provide a
qualitatively different, complementary test.
We show that when the long-lived charged particle producing a
displaced-lepton signal at the High-Luminosity LHC is also the mediator
responsible for Majorana neutrino masses---the \textit{shared-coupling
condition}---the Casas--Ibarra parametrization determines the lepton
flavor ratio $N_e/N_\mu$ of time-delayed leptons in terms of
Pontecorvo--Maki--Nakagawa--Sakata (PMNS) mixing parameters alone.
Within the resulting \textit{shared-coupling class} of models the ratio is
fixed by oscillation data, independent of all beyond-Standard-Model mass
and coupling parameters up to calculable corrections bounded by
charged-lepton-flavor-violation data, yielding $N_e/N_\mu = 0.140$ for the
Normal and $2.11$ for the Inverted Hierarchy---a separation of more than
an order of magnitude between the two predictions.
This prediction holds identically across the Scotogenic model, the
Type-III seesaw, the inverse and linear seesaw, and lepton-portal dark
matter, and is stable against oscillation-parameter uncertainties and
detector-efficiency variations, which cancel in the ratio.
A single efficiency-corrected decision criterion identifies the
hierarchy, and explicit signal-yield estimates combined with an exact
Poisson test show that $\mathcal{O}(25)$
displaced-lepton events suffice for a $3\sigma$ discrimination with the
timing infrastructure already under construction at CMS and ATLAS---so
that a positive long-lived-particle signal would simultaneously resolve
the mass ordering, with no dependence on $\delta_{CP}$.

\vspace{0.5em}

\end{abstract}

\maketitle

%=============================================================================
\section{Introduction}
\label{sec:intro}
%=============================================================================

The neutrino mass ordering---Normal Hierarchy (NH: $m_3 \gg m_1, m_2$) or
Inverted Hierarchy (IH: $m_1 \approx m_2 \gg m_3$)---remains one of the
central open questions in particle
physics~\cite{deGouvea2016,Esteban2020nufit,Atre:2009rg,Cai:2017mow}.
The joint T2K and NO$\nu$A analysis~\cite{T2K:2025wet} finds no strong
preference for either mass ordering, and definitive resolution remains
elusive owing to persistent $\delta_{CP}$--hierarchy degeneracies:
the oscillation probability $P(\nu_\mu \to \nu_e)$ in matter is entangled
with the unknown CP phase $\delta_{CP}$, forcing each long-baseline
experiment to measure both quantities simultaneously, at significant cost
in sensitivity.
JUNO~\cite{JUNO:2024jaw,JUNO:2025gmd} has recently begun taking data and
aims to determine the ordering via reactor antineutrino vacuum
oscillations, independently of $\delta_{CP}$, within ${\sim}6.5$ years;
definitive confirmation from long-baseline experiments will require
DUNE~\cite{DUNE2022} and Hyper-K~\cite{HyperK2018} on a decade-scale
timescale.

The HL-LHC precision timing detectors---the MIP Timing Detector (MTD) at
CMS~\cite{CMS_MTD} and the High-Granularity Timing Detector (HGTD) at
ATLAS~\cite{ATLAS_HGTD}---offer a structurally different probe.
Achieving ${\sim}30~\mathrm{ps}$ time resolution, they isolate long-lived
particle (LLP) decay products via a time-delay requirement
$\dt > 200~\mathrm{ps}$, suppressing prompt Standard Model backgrounds to
levels estimable with data-driven timing sidebands~\cite{Liu2019timing}.
In this Letter we show that, when the LLP is the mediator responsible for
Majorana neutrino masses, the flavor ratio $\Ne/\Nmu$ of time-delayed
leptons is a $\delta_{CP}$-independent hierarchy discriminant fixed
entirely by Pontecorvo--Maki--Nakagawa--Sakata (PMNS) mixing angles and
neutrino mass eigenvalues---a
connection not identified in prior
work~\cite{SUSY-2020-08,CMS-EXO-23-016,Alimena2020LLP}.
We refer to this measurement strategy as \textit{flavor tomography}: the
flavor composition of the LLP decay products directly images the
underlying neutrino mass spectrum, so that counting displaced electrons
and muons reconstructs information about the mass ordering.

Using collider lepton-flavor counting to probe the neutrino mass
hierarchy has a history.
Earlier work showed that branching ratios of doubly-charged Higgs bosons
in the Type-II seesaw~\cite{Garayoa:2007fw,Kadastik:2007yd},
triplet-fermion decays in the Type-III seesaw~\cite{Eboli:2011ia},
four-lepton signatures at future colliders~\cite{Mandal:2022ysp}, and
heavy neutral lepton (HNL) decays at $Z$ factories~\cite{Liu:2026nmo}
all carry an imprint of the mass ordering.
Three limitations are shared by these approaches.
\textit{First}, they depend on specific model architectures---Type-II or
Type-III seesaw, or two-HNL Type-I---and the discriminating power is not
portable to other neutrino mass models.
\textit{Second}, none exploits precision timing: they rely on prompt or
mildly displaced signatures and require conventional detector simulation
to relate observed event rates to theoretical predictions.
\textit{Third}, and most importantly, the Dirac CP phase $\delta_{CP}$
contaminates the hierarchy prediction, broadening the predicted NH and IH
flavor regions until they overlap; a quantitative comparison with the
two-HNL analysis of Ref.~\cite{Liu:2026nmo} is given in
Sec.~\ref{sec:comparison}, after our central result has been derived.

The present work addresses all three limitations for the class of models
defined in Sec.~\ref{sec:framework}, in which the LLP decay coupling is
tied to the neutrino mass matrix by the Casas--Ibarra
identity~\cite{CasasIbarra2001,Atre:2009rg,Cai:2017mow,Herrero-Garcia:2025aox,Herrero-Brocal:2025zpb}:
within this class, every beyond-Standard-Model (BSM) free parameter drops
out of $\Ne/\Nmu$, and the resulting prediction is a
$\delta_{CP}$-independent hierarchy discriminant that requires no
model-specific detector simulation.
The remainder of this Letter is organized as follows.
Section~\ref{sec:framework} defines the shared-coupling class, derives
the universal flavor-ratio prediction, explains its analytic origin, and
contrasts it with previous flavor-counting proposals.
Section~\ref{sec:models} verifies the prediction explicitly in four
representative BSM models.
Section~\ref{sec:robustness} establishes its robustness against the CP
phase, oscillation-parameter uncertainties, the absolute neutrino mass
scale, the BSM spectrum, and detector efficiencies.
Section~\ref{sec:hllhc} presents the experimental decision criterion and
the statistical reach at the HL-LHC, and Sec.~\ref{sec:summary}
summarizes.
Three appendices collect the dependence on the lightest neutrino mass
(Appendix~A), numerical validation scans
(Appendix~B), and the binomial statistical test
(Appendix~C).

%=============================================================================
\section{The shared-coupling class and the universal flavor ratio}
\label{sec:framework}
%=============================================================================

%-----------------------------------------------------------------------------
\subsection{Definition and assumptions}
\label{sec:definition}
%-----------------------------------------------------------------------------

We consider a BSM model containing a charged LLP $X^\pm$ that decays as
$X^\pm \to \ell^\pm_\alpha + \text{invisible}$ via a Yukawa coupling
matrix $\lambda_{a\alpha}$, where $a$ labels the members of the heavy
mediator multiplet $\{X_a\}$ and $\alpha = e, \mu, \tau$ the lepton
flavor.
The central hypothesis of this Letter is stated as follows.

\medskip
\noindent\textbf{Definition (shared-coupling class).}
A BSM model belongs to the \textit{shared-coupling class} if it satisfies
the following three conditions:
\begin{itemize}
  \item[(i)] \textit{Shared-coupling condition}: the Yukawa matrix
    $\lambda_{a\alpha}$ governing the LLP decay is the same matrix that
    generates the Majorana neutrino mass matrix through a seesaw or
    radiative mechanism.
  \item[(ii)] \textit{Quasi-degenerate mediators}: the heavy mediator
    masses satisfy $M_a \approx M$, so that loop functions and mass
    factors cancel in the ratio of branching ratios.
  \item[(iii)] \textit{Approximately real $R$}: the Casas--Ibarra matrix
    $R$ is close to real orthogonal ($R^T R = \mathbf{1}$).
\end{itemize}
\medskip

Throughout this Letter, the term \textit{shared-coupling condition}
refers to condition (i) alone, while the \textit{shared-coupling class}
denotes the set of models satisfying all three conditions (i)--(iii).
The three conditions play distinct roles: condition (i) projects the LLP
decay coupling onto the PMNS matrix via the Casas--Ibarra parametrization;
conditions (ii) and (iii) guarantee that the remaining BSM parameters---%
mediator masses and $R$-matrix angles---cancel in the flavor ratio.
All three are required for the universal prediction derived below.
Conditions (ii) and (iii) are approximate rather than exact in realistic
models; the corrections they induce are quantified in
Sec.~\ref{sec:robustness} and Appendix~B, and are
bounded by charged-lepton-flavor-violation (cLFV) data to the
$\mathcal{O}(10\text{--}30\%)$ level, far below the hierarchy separation
established below.

%-----------------------------------------------------------------------------
\subsection{The universal flavor-ratio prediction}
\label{sec:prediction}
%-----------------------------------------------------------------------------

\noindent\textbf{Proposition.}
For any model in the shared-coupling class, the inclusive flavor ratio of
LLP decays into electrons and muons is
\begin{equation}
  \frac{\Ne}{\Nmu}\bigg|_{\rm theory}
  = \frac{\sum_i |U_{ei}|^2\, m_i}{\sum_i |U_{\mu i}|^2\, m_i},
  \label{eq:universal_ratio}
\end{equation}
a function of PMNS matrix elements $U_{\alpha i}$ and neutrino mass
eigenvalues $m_i$ alone, with no dependence on any BSM mass or coupling
parameter.

\medskip
The proof is short.
If the matrix $\lambda_{a\alpha}$ generates Majorana neutrino masses,
the Casas--Ibarra parametrization~\cite{CasasIbarra2001,Herrero-Garcia:2025aox}
gives $\lambda_{a\alpha} \sim \sum_i R_{ai}\sqrt{m_i}\,U^*_{\alpha i}/v$.
Summing inclusively over the quasi-degenerate mediator multiplet
$\{X_a\}$, the partial width into flavor $\alpha$ obeys
\begin{equation}
  \sum_a \mathrm{BR}(X_a \to \ell^\pm_\alpha)
  \;\propto\;
  \sum_a |\lambda_{a\alpha}|^2
  \;\propto\;
  \sum_i |U_{\alpha i}|^2\, m_i,
  \label{eq:universal_BR}
\end{equation}
where the second proportionality uses the real orthogonality
$R^T R = \mathbf{1}$
of condition (iii) and the equal mediator masses of condition (ii),
and all flavor-independent factors drop out of the ratio below.
Taking the ratio of the $e$ and $\mu$ channels yields
Eq.~(\ref{eq:universal_ratio}).
The Proposition is exact when conditions (ii) and (iii) hold exactly;
when they hold only approximately, Eq.~(\ref{eq:universal_ratio})
acquires the calculable corrections quantified in
Sec.~\ref{sec:robustness} and Appendix~B.

Evaluating Eq.~(\ref{eq:universal_ratio}) at the \NuFit\ best-fit
oscillation parameters~\cite{Esteban2020nufit} with
$m_{\rm lightest} = 0$ gives
\begin{equation}
  \frac{\Ne}{\Nmu}\bigg|_{\rm theory}
  = \begin{cases}
      0.140 & (\NH), \\
      2.11 & (\IH).
    \end{cases}
  \label{eq:benchmark_values}
\end{equation}
It is convenient to define the \textit{hierarchy discrimination ratio}
\begin{equation}
  \Rdisc \;\equiv\;
  \frac{(\Ne/\Nmu)_{\rm IH}}{(\Ne/\Nmu)_{\rm NH}}
  \;=\; \frac{2.11}{0.140} \;=\; 15.1
  \label{eq:Rdisc}
\end{equation}
at this benchmark point; $\Rdisc$ quantifies the separation between the
two hierarchy predictions and is used throughout the remainder of this
Letter.
Table~\ref{tab:inputs} lists the full set of input parameters and the
resulting flavor-weighted mass sums, ensuring that
Eqs.~(\ref{eq:universal_ratio})--(\ref{eq:Rdisc}) are transparently
reproducible.
Equations~(\ref{eq:universal_ratio})--(\ref{eq:Rdisc}) constitute the
central result of this Letter.

\begin{table}[h]
\centering
\caption{NuFit~6.1 input parameters~\cite{Esteban2020nufit} (IC23
without SK, $m_{\rm lightest}=0$) and the resulting flavor-weighted
mass sums $\mathcal{A}_\alpha \equiv \sum_i |U_{\alpha i}|^2 m_i$
(meV) that determine the $\Ne/\Nmu$ prediction of
Eq.~(\ref{eq:universal_ratio}).
Each hierarchy is evaluated at its own NuFit~6.1 best-fit parameter set;
in particular, $\sin^2\theta_{23} = 0.470$ (NH) vs.\ $0.553$ (IH),
which is the dominant source of the difference in $\mathcal{A}_\mu$.}
\label{tab:inputs}
\begin{tabular}{lcc}
\hline
Quantity & NH & IH \\
\hline
$\sin^2\theta_{12}$ & $0.307$ & $0.307$ \\
$\sin^2\theta_{13}$ & $0.02249$ & $0.02237$ \\
$\sin^2\theta_{23}$ & $0.470$ & $0.553$ \\
$\delta_{CP}$ & $232^\circ$ & $197^\circ$ \\
$m_1, m_2, m_3$ [meV] & $0,\;8.6,\;50.1$ & $48.6,\;49.4,\;0$ \\
\hline
$\mathcal{A}_e$ [meV] & $3.71$ & $47.75$ \\
$\mathcal{A}_\mu$ [meV] & $26.57$ & $22.63$ \\
$\Ne/\Nmu$ & $\mathbf{0.140}$ & $\mathbf{2.11}$ \\
\hline
\end{tabular}
\end{table}

%-----------------------------------------------------------------------------
\subsection{Analytic origin of the NH--IH separation}
\label{sec:analytic}
%-----------------------------------------------------------------------------

The benchmark values in Eq.~(\ref{eq:benchmark_values}) admit simple
analytic approximations that expose which oscillation parameters control
the separation.
The two expressions derived in this subsection are leading-order
approximations to the exact result Eq.~(\ref{eq:universal_ratio}); they
are not independent predictions, but serve to trace the value of $\Rdisc$
to two well-measured PMNS angles.

For NH, with $m_1 \approx 0$ and $m_3/m_2 \approx 5.8$, the heaviest
eigenstate $\nu_3$ dominates Eq.~(\ref{eq:universal_ratio}).
Because $|U_{e3}|^2 = \sin^2\theta_{13}$ while
$|U_{\mu 3}|^2 \approx \cos^2\theta_{13}/2$ at maximal $\theta_{23}$,
the ratio is suppressed by the smallness of reactor mixing:
\begin{equation}
  \frac{\Ne}{\Nmu}\bigg|_{\rm NH}
  \approx
  \frac{\sin^2\theta_{13}}{\cos^2\theta_{13}/2}
  = 2\tan^2\theta_{13}
  \approx 0.046.
  \label{eq:NH_approx}
\end{equation}
The leading-$\nu_3$ approximation gives $0.046$; the exact \NuFit\
value $0.140$ includes the sub-leading $\nu_2$ contribution
($m_2 \approx 8.6~\mathrm{meV}$), confirming Eq.~(\ref{eq:NH_approx}) as
an analytic lower bound.

For IH, with $m_3 \approx 0$, the muon-rich $\nu_3$ is decoupled, leaving
the quasi-degenerate pair $\nu_1, \nu_2$.
Their summed entries satisfy
$|U_{e1}|^2 + |U_{e2}|^2 = 1 - \sin^2\theta_{13}$ and
$|U_{\mu 1}|^2 + |U_{\mu 2}|^2 \approx \cos^2\theta_{13}/2$
(exact at $\theta_{23} = 45^\circ$),
yielding
\begin{equation}
  \frac{\Ne}{\Nmu}\bigg|_{\rm IH}
  \approx
  \frac{1 - \sin^2\theta_{13}}{\cos^2\theta_{13}/2}
  \approx 1.91 \quad (\theta_{23} = 45^\circ).
  \label{eq:IH_approx}
\end{equation}
Evaluating instead at the \NuFit\ IH best-fit
$\sin^2\theta_{23}^{\rm IH} = 0.553$ ($\theta_{23}^{\rm IH} = 48.2^\circ$),
which enhances the electron-flavor content of the surviving $\nu_{1,2}$
pair, raises the degenerate-pair estimate to $2.13$; the full calculation
(retaining the small $m_1 \neq m_2$ splitting and $\delta_{CP}$)
gives the exact value $2.11$.
The discrimination ratio $\Rdisc = 15.1$ is thus set by the same two
angles---$\theta_{13}$ and $\theta_{23}$---that reactor and atmospheric
experiments have measured most precisely.

%-----------------------------------------------------------------------------
\subsection{Relation to previous flavor-counting approaches}
\label{sec:comparison}
%-----------------------------------------------------------------------------

With the central result in hand, the comparison with earlier
flavor-counting proposals can be made quantitative, using the two-HNL
analysis at $Z$ factories~\cite{Liu:2026nmo} as the most developed
representative.
As explicitly shown in Ref.~\cite{Liu:2026nmo}, the NH prediction for
$U^2_{eN}/U^2_{\rm tot}$ spans the wide range
$(3.1\times10^{-3},\,0.14)$ when the oscillation parameters are scanned
over their $3\sigma$ ranges, contracting only once $\delta_{CP}$ is
pinned down externally by dedicated oscillation experiments.
Because the predicted NH and IH regions overlap intrinsically under this
scan---with the breadth of the NH region driven in large part by the
unconstrained $\delta_{CP}$---the discrimination power (the
correct-classification probability, denoted $K$ in
Ref.~\cite{Liu:2026nmo}) is bounded by a
hard ceiling of ${\lesssim}90\%$ even in the limit of infinite
statistics, independently of collider luminosity.
By contrast, the flavor-tomography ratio $\Ne/\Nmu$ of
Eq.~(\ref{eq:universal_ratio}) varies by less than $3\%$ across all
values of $\delta_{CP}$ for real Casas--Ibarra $R$ matrices
(Sec.~\ref{sec:robustness}), so the NH and IH predictions of
Eq.~(\ref{eq:benchmark_values}) are well separated for every value of
$\delta_{CP}$, with no overlap within the $3\sigma$ oscillation band.

This sharpness has a structural origin that distinguishes the present
observable from the two-HNL analysis at a deeper level than statistics.
In the minimal seesaw with two heavy states, the flavor ratios
$U^2_{\alpha N}/U^2_{\rm tot}$ retain an explicit dependence on the
complex Casas--Ibarra angle through
$x_\omega \equiv e^{\,\mathrm{Im}\,\omega}$
[Ref.~\cite{Liu:2026nmo}, Eq.~(1)], which does not cancel in the ratio
for only two heavy states and broadens the allowed flavor region beyond
the irreducible oscillation-parameter spread.
The shared-coupling observable removes this dependence at its root:
summing inclusively over a quasi-degenerate mediator multiplet of three
or more states with real orthogonal $R$ collapses
$\sum_a |\lambda_{a\alpha}|^2$ to $\sum_i |U_{\alpha i}|^2 m_i$
[Eq.~(\ref{eq:universal_BR})], so $\Ne/\Nmu$ reduces to a function of
PMNS data alone, with complex-$R$ corrections bounded by cLFV data to
$\mathcal{O}(10\text{--}30\%)$ (Sec.~\ref{sec:robustness} and
Appendix~B).
The present prediction is therefore sharp precisely where the two-HNL
flavor ratio is intrinsically broad.

%=============================================================================
\section{Model realizations}
\label{sec:models}
%=============================================================================

Table~\ref{tab:universality} lists four representative BSM models
belonging to the shared-coupling class; in each case the Yukawa structure
reduces to Eq.~(\ref{eq:universal_BR}) through the Casas--Ibarra
identity, so the prediction of Eq.~(\ref{eq:universal_ratio}) holds in
all four without any model-specific input.
We first work through the Scotogenic model in detail and then verify the
remaining three.

\begin{table}[h]
\centering
\caption{Representative BSM models belonging to the shared-coupling
class (conditions (i)--(iii) of Sec.~\ref{sec:definition}).
The $\Ne/\Nmu$ prediction of Eq.~(\ref{eq:universal_ratio}) is identical
across all four for a given hierarchy; the Scotogenic decay is inclusive
over the three $N_i$.}
\label{tab:universality}
\footnotesize
\setlength{\tabcolsep}{2pt}
\begin{tabular}{llcc}
\hline
  Model & LLP decay & $\nu$ mass & $M_{\rm LLP}$ [GeV] \\
  \hline
  Scotogenic~\cite{Ma2006,Heeck:2022rep}
    & $\etapm \!\to\! \ell^\pm N_i$ & 1-loop  & $100$--$800$ \\
  Type-III seesaw~\cite{Foot1989,Franceschini2008}
    & $\Sigma^\pm \!\to\! \ell^\pm Z/h$ & tree    & $100$--$1000$ \\
  Inv./linear seesaw~\cite{Mohapatra1986,Malinsky2005}
    & $N \!\to\! \ell^\pm W^\mp$        & tree    & $100$--$500$ \\
  Lepton-portal DM~\cite{Batell2017}
    & $\phi^\pm \!\to\! \ell^\pm \chi$  & 1-loop  & $100$--$600$ \\
    \hline
\end{tabular}
\end{table}

\textit{Scotogenic model.}
The Scotogenic model~\cite{Ma2006} extends the SM by a $\mathbb{Z}_2$-odd
inert scalar doublet $\Phi = (\etapm, \tilde{\eta}^0)$ and three
right-handed neutrinos $N_i$.
Neutrino masses arise at one loop via the same dark Yukawa coupling
$y_{i\alpha}$ that also drives the two-body decays
$\etapm \to \ell^\pm_\alpha + N_i$ (one channel per $N_i$):
\begin{equation}
  (\mathcal{M}_\nu)_{\alpha\beta}
  = \sum_i \frac{y_{i\alpha}\, y_{i\beta}\, M_{N_i}}{16\pi^2}\,
    \mathcal{F}(M_{N_i},\, m_\eta),
  \label{eq:Mnu}
\end{equation}
where $\mathcal{F}$ is the one-loop integral function~\cite{Ma2006}.
The decay coupling and the mass-generating coupling are one and the same
matrix, so the shared-coupling condition (i) holds by construction.
With the Casas--Ibarra parametrization and degenerate
$M_{N_i} \equiv M_N$ [condition (ii)], $\mathcal{F}$ is common to all
channels and the total partial width into flavor $\alpha$, summed
inclusively over all $N_i$ final states, reduces to
$\Gamma(\etapm \to \ell^\pm_\alpha + \text{any } N_i) \propto
\sum_i |y_{i\alpha}|^2 \propto \sum_i |U_{\alpha i}|^2 m_i$, recovering
Eq.~(\ref{eq:universal_BR}) exactly.
The signal sample therefore consists of all $\etapm$ decays to a charged
lepton plus any invisible $N_i$, detected inclusively via the timing cut;
the individual $N_i$ are not required to be resolved.
Inserting the \NuFit\ best-fit mass eigenvalues of Table~\ref{tab:inputs}
and retaining the $e$ and $\mu$ channels (normalized within the $e/\mu$
subset, with the $\tau$ channel treated as signal loss) gives
\begin{equation}
  \begin{aligned}
    \NH:\quad & \mathrm{BR}(e) = 12.2\%,\quad \mathrm{BR}(\mu) = 87.8\%, \\
    \IH:\quad & \mathrm{BR}(e) = 67.9\%,\quad \mathrm{BR}(\mu) = 32.1\%.
  \end{aligned}
  \label{eq:BR}
\end{equation}
The $\tau$ channel (${\approx}48\%$ for NH, ${\approx}28\%$ for IH of the
total width) contributes less than $1\%$ to $\Ne/\Nmu$ after the
$\dt > 200~\mathrm{ps}$ selection, owing to leptonic sub-branching and
timing suppression of secondary vertices
(Appendix~B).

The other three models in Table~\ref{tab:universality} satisfy the
shared-coupling condition through related but structurally distinct
mechanisms; we now verify each explicitly.

\textit{Type-III seesaw.}
The triplet fermion $\Sigma^\pm$ has mass matrix
$\mathcal{M}_\nu = -v^2 Y^T M_\Sigma^{-1} Y / 2$, identical in structure
to the Type-I seesaw, so the Casas--Ibarra parametrization gives
$|Y_{i\alpha}|^2 \propto |U_{\alpha i}|^2 m_i / M_{\Sigma_i}$.
The charged decays $\Sigma^\pm \to \ell^\pm_\alpha + Z/h$ proceed via
lepton--triplet mixing of order $\lambda v/M$~\cite{Franceschini2008},
with both channels proportional to $\sum_i |Y_{i\alpha}|^2 M$ and no
additional flavor-dependent prefactor.
For degenerate $M_{\Sigma_i} \equiv M_\Sigma$ [condition (ii)],
Eq.~(\ref{eq:universal_BR}) is recovered exactly for the combined $Z/h$
mode throughout the HL-LHC accessible range
$M_\Sigma \gtrsim 150~\mathrm{GeV}$~\cite{Das:2020uer}.

\textit{Inverse and linear seesaw.}
Both variants give $\mathcal{M}_\nu \propto Y_D \mu Y_D^T$, and the
Casas--Ibarra parametrization yields
$\sum_i |(Y_D)_{i\alpha}|^2 \propto \sum_i |U_{\alpha i}|^2 m_i$
for real $R$~\cite{Mohapatra1986,Malinsky2005}.
The HNL decay $N \to \ell^\pm_\alpha + W^\mp$ proceeds via active--sterile
mixing $|\Theta_{\alpha i}|^2 \propto |U_{\alpha i}|^2 m_i / M_{N_i}^2$;
the $M_{N_i}^2$ factor cancels in the ratio of branching ratios for
degenerate $M_{N_i} \equiv M_N$, recovering Eq.~(\ref{eq:universal_BR})
exactly.
The degeneracy required by condition (ii) is naturally realized in the
HL-LHC parameter space ($M_N \sim 100$--$500~\mathrm{GeV}$,
$\mu \ll M_N$), where the pseudo-Dirac splitting
$\Delta M \sim \mu \ll M_N$ keeps both eigenstates nearly degenerate and
the deviation from Eq.~(\ref{eq:universal_ratio}) is of order
$\Delta M/M_N \ll 1$.

\textit{Lepton-portal dark matter.}
The charged mediator $\phi^\pm$ couples to SM leptons and a DM fermion
$\chi$ via
$\mathcal{L} \supset \lambda_{i\alpha}\,\bar{\ell}_\alpha\,\phi^-\,\chi_i
+ \mathrm{h.c.}$
Neutrino masses arise radiatively through the same $\lambda_{i\alpha}$
at one loop, with the same mass-matrix structure as the Scotogenic
model~\cite{Batell2017,Ma2006}.
The shared-coupling condition therefore holds for lepton-portal DM by
the same argument as the Scotogenic case, and Eq.~(\ref{eq:universal_BR})
follows directly.

%=============================================================================
\section{Robustness of the prediction}
\label{sec:robustness}
%=============================================================================

Six potential sources of uncertainty---the CP phase, the measured
oscillation parameters, the absolute neutrino mass scale, the BSM
mediator spectrum, the $R$-matrix freedom, and detector efficiencies---%
leave the hierarchy discrimination intact.
We address each in turn.

\textit{(i) The Dirac CP phase.}
For real Casas--Ibarra $R$ [condition (iii)], $\delta_{CP}$ cancels
exactly in $\Ne/\Nmu$: the observable depends on
$|U_{\alpha i}|^2 = U_{\alpha i}U_{\alpha i}^*$, which is manifestly
phase-independent for fixed mixing angles and masses.
The small variation seen in a numerical scan ($<3\%$ for NH, $<0.5\%$ for
IH) arises because NuFit's best-fit parameters
$(\theta_{ij}, \delta_{CP})$ are jointly determined from global data:
scanning $\delta_{CP}$ at the corresponding best-fit $\theta_{ij}$ values
introduces an indirect correlation, not a genuine phase dependence.
For fixed mixing angles, the independence is exact.
This is in sharp contrast to the matter-modified probability
$P(\nu_\mu \to \nu_e)$ that long-baseline experiments rely on, where the
$\delta_{CP}$--hierarchy degeneracy produces $\mathcal{O}(1)$
shifts~\cite{deGouvea2016}; the present observable is robust against an
unknown CP phase within the shared-coupling class.

\begin{figure}[t]
\centering
\includegraphics[width=0.9\columnwidth]{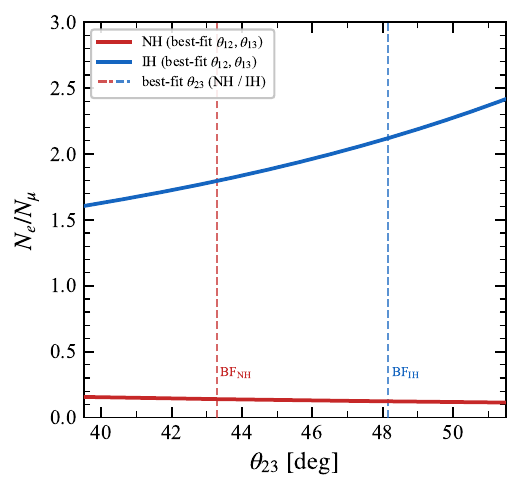}
\caption{Predicted $\Ne/\Nmu$ for NH (red) and IH (blue) vs.\
$\theta_{23}$, at the \NuFit\ best-fit values of $\theta_{12}$ and
$\theta_{13}$~\cite{Esteban2020nufit}.
Vertical dashed lines mark the \NuFit\ best-fit $\theta_{23}$ for NH
($43.3^\circ$) and IH ($48.2^\circ$).
Across the full $3\sigma$ range of $\theta_{23}$ the two curves remain
separated by more than the global worst-case value $\Rdisc = 8.85$,
which is attained only in the full three-angle $3\sigma$ scan described
in the text; improved oscillation measurements will
only sharpen the discrimination.}
\label{fig:oscparams}
\end{figure}

\textit{(ii) Oscillation-parameter uncertainties.}
Scanning $5 \times 10^4$ points uniformly over the \NuFit\ $3\sigma$
ranges ($\theta_{12} \in [32.54^\circ, 35.03^\circ]$,
$\theta_{13} \in [8.27^\circ, 8.95^\circ]$,
$\theta_{23} \in [41.11^\circ, 50.02^\circ]$)~\cite{Esteban2020nufit}
at $m_{\rm lightest} = 0$ yields
NH: $\Ne/\Nmu \in [0.099, 0.181]$ and IH: $[1.602, 2.490]$.
Even for the most unfavorable parameter combination---the IH minimum
($1.602$) against the NH maximum ($0.181$)---the discrimination ratio
is still as large as $\Rdisc = 8.85$ (smaller than earlier estimates that used a
single $\theta_{23}$ value for both hierarchies, because here each
hierarchy is evaluated at its own NuFit~6.1 best-fit $\theta_{23}$); as
oscillation experiments tighten $\theta_{23}$, this gap widens rather
than narrows (Fig.~\ref{fig:oscparams}).

\textit{(iii) Lightest neutrino mass.}
The full dependence of $\Ne/\Nmu$ on $m_{\rm lightest}$ is given in
Appendix~A.
Within the conservative Planck 2018 bound~\cite{Planck2018}
($\sum_i m_i < 120~\mathrm{meV}$; the DESI BAO
constraint~\cite{DESI:2024mwx} is tighter but
cosmology-combination-dependent, so we adopt Planck 2018 as the
conservative baseline), the IH prediction falls no lower than
${\approx}1.58$ (reached at its Planck-allowed maximum
$m_{\rm lightest} \approx 15~\mathrm{meV}$) and the NH prediction rises
at most to $0.71$ (at $m_{\rm lightest} \approx 30~\mathrm{meV}$); the
discrimination ratio therefore satisfies
$\Rdisc \geq 2.2$ ($\approx 1.58/0.71$) throughout the Planck-allowed
range.
In practice, current cosmological fits and the direct upper limit
$m_\nu < 0.45~\mathrm{eV}$ from KATRIN~\cite{KATRIN2025} are consistent
with $m_{\rm lightest} \lesssim 15~\mathrm{meV}$, where the
discrimination ratio satisfies $\Rdisc > 2.9$.

\textit{(iv) Non-degenerate mediator masses.}
A scan over non-degenerate right-handed-neutrino spectra in the
Scotogenic model, with mass splittings up to $50\%$, shows that the
discrimination ratio satisfies $\Rdisc > 9$ in all cases
(Appendix~B).

\textit{(v) $R$-matrix freedom.}
For real orthogonal $R$, $\sum_i |y_{i\alpha}|^2$ is independent of $R$
and Eq.~(\ref{eq:universal_ratio}) holds exactly.
In the parameter space compatible with the MEG~II
bound~\cite{MEGII2025} ($\mathrm{BR}(\mu\to e\gamma) < 1.5\times10^{-13}$),
a scan over $10^5$ complex-$R$ matrices shows deviations of at most
$\mathcal{O}(10\text{--}30\%)$ in $\Ne/\Nmu$---far below the factor
$\Rdisc = 15.1$ separating the NH and IH predictions
(scan details in Appendix~B).

\begin{figure}[t]
\centering
\includegraphics[width=0.9\columnwidth]{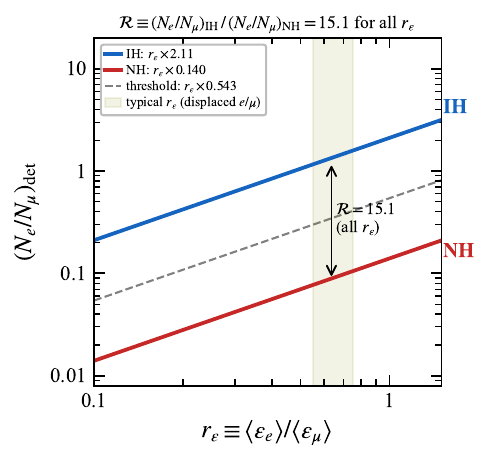}
\caption{Detector-level ratio $(N_e/N_\mu)_{\rm det}$ vs.\
$r_\varepsilon \equiv \langle\varepsilon_e\rangle/\langle\varepsilon_\mu\rangle$
for NH (red, theory value $0.140$) and IH (blue, theory value $2.11$).
Both lines have slope unity on the log--log scale, reflecting the exact
cancellation of $r_\varepsilon$ in Eq.~(\ref{eq:det_ratio}), with a
constant separation of $\Rdisc = 15.1$ for any $r_\varepsilon > 0$.
The dashed gray line shows the hierarchy decision threshold
$r_\varepsilon \times 0.543$ derived in Sec.~\ref{sec:hllhc}; it lies between
the two lines for all $r_\varepsilon > 0$.
Olive band: typical displaced-track regime
$r_\varepsilon \approx 0.55$--$0.75$ at ATLAS and CMS.
Theory inputs: \NuFit\ best-fit~\cite{Esteban2020nufit},
$m_{\rm lightest} = 0$, each hierarchy evaluated at its own best-fit
parameters (Table~\ref{tab:inputs}).}
\label{fig:reps}
\end{figure}

\textit{(vi) Detector efficiency.}
At detector level the measured ratio differs from the theoretical
prediction by the relative efficiency
$r_\varepsilon \equiv \langle\varepsilon_e\rangle/\langle\varepsilon_\mu\rangle$:
\begin{equation}
  (\Ne/\Nmu)_{\rm det} = r_\varepsilon \cdot (\Ne/\Nmu)_{\rm theory}.
  \label{eq:det_ratio}
\end{equation}
Because $r_\varepsilon$ enters as a common multiplicative factor, it
cancels identically in the NH--IH separation:
$(N_e/N_\mu)^{\rm IH}_{\rm det} / (N_e/N_\mu)^{\rm NH}_{\rm det}
= \Rdisc = 15.1$ for any $r_\varepsilon > 0$.
Figure~\ref{fig:reps} illustrates this: on a log--log scale the NH and IH
lines are parallel with slope unity, separated by $\log \Rdisc$, with the
hierarchy decision threshold $r_\varepsilon \times 0.543$, derived in
Sec.~\ref{sec:hllhc}, between them for all $r_\varepsilon > 0$.
The value of $r_\varepsilon$ can be measured from a displaced-lepton
control sample; suitable candidates include displaced
$J/\psi \to \ell^+\ell^-$ decays reconstructed in the timing layer, which
share the same lepton-identification and timing-selection chain as the
signal without requiring any BSM model input~\cite{CMS_MTD}.
The same control sample constrains the flavor asymmetry of any residual
background, as discussed in Sec.~\ref{sec:hllhc}.
Applying Eq.~(\ref{eq:det_ratio}) therefore does not rely on
model-specific full detector reinterpretation of the BSM signal.

%=============================================================================
\section{Hierarchy discrimination at the HL-LHC}
\label{sec:hllhc}
%=============================================================================

%-----------------------------------------------------------------------------
\subsection{Decision criterion and Majorana test}
\label{sec:criterion}
%-----------------------------------------------------------------------------

The $3\sigma$ oscillation scan of Sec.~\ref{sec:robustness} establishes a
clean gap between the NH ceiling ($\Ne/\Nmu \leq 0.181$) and the IH floor
($\Ne/\Nmu \geq 1.602$).
Correcting the observed ratio for efficiency via
Eq.~(\ref{eq:det_ratio}) gives the hierarchy decision criterion within
the shared-coupling class:
\begin{equation}
  \boxed{
    \frac{(N_e/N_\mu)^{\rm obs}}{r_\varepsilon}
    \begin{cases}
      < 0.543 & \Rightarrow \text{Normal Hierarchy,} \\
      > 0.543 & \Rightarrow \text{Inverted Hierarchy,}
    \end{cases}
  }
  \label{eq:criterion}
\end{equation}
where the threshold $0.543$ is the geometric midpoint
$\sqrt{(N_e/N_\mu)^{\rm NH} \cdot (N_e/N_\mu)^{\rm IH}}
= \sqrt{0.140\times2.11} = 0.543$
evaluated at $m_{\rm lightest} = 0$; this is a \emph{benchmark} value
appropriate for the cosmologically preferred regime.
The criterion is valid for $m_{\rm lightest} \lesssim 15~\mathrm{meV}$,
consistent with current cosmological
constraints~\cite{Planck2018,DESI:2024mwx}; for larger masses the
$m_{\rm lightest}$-dependent threshold of Appendix~A
should be substituted, though
$m_{\rm lightest} \gtrsim 15~\mathrm{meV}$ is disfavored and the
discrimination ratio remains $\Rdisc \geq 2.2$ throughout the
Planck-allowed range.

The flavor-tomography observable is also inextricably tied to the
Majorana nature of neutrino masses: if neutrinos are Dirac particles, the
Casas--Ibarra relation no longer constrains the coupling to the PMNS
matrix, and $\Ne/\Nmu$ becomes a free parameter unconstrained by
oscillation data.
Agreement of the observed ratio with Eq.~(\ref{eq:universal_ratio})
would provide evidence that the LLP coupling satisfies the
shared-coupling condition consistent with Majorana mass generation.
We note that this tests the coupling structure rather than the Majorana
mass term directly; a definitive Majorana determination additionally
requires a lepton-number-violating observable such as same-sign dilepton
production.

%-----------------------------------------------------------------------------
\subsection{Signal yields}
\label{sec:yields}
%-----------------------------------------------------------------------------

We estimate the signal yield for three benchmark points spanning the LLP
mass range of Table~\ref{tab:universality}, using the Scotogenic model as
a concrete example.
The $pp \to \etamp$ production cross section at
$\sqrt{s} = 14~\mathrm{TeV}$ is computed at LO with
\textsc{MadGraph5\_aMC@NLO}~3.5.13~\cite{Alwall2014} using the
\texttt{InertDoublet\_UFO} model~\cite{Goudelis:2013uca,Belanger:2015kga}
with \texttt{NNPDF2.3LO} PDFs~\cite{Ball:2012cx}, giving
$\sigma_{\rm LO} = 3.66~\mathrm{fb}$ ($M_{\rm LLP} = 300~\mathrm{GeV}$),
$1.09~\mathrm{fb}$ ($400~\mathrm{GeV}$), and $0.165~\mathrm{fb}$
($600~\mathrm{GeV}$).
Applying an NLO $K$-factor $K \approx 1.3$, consistent with NLO QCD
calculations for color-singlet scalar pair production via Drell--Yan at
this mass range~\cite{Ghosh:2021noq,Ahmed:2025jjl}, gives
$\sigma_{\rm NLO} = 4.76$, $1.41$, and $0.214~\mathrm{fb}$, respectively.
With $\mathcal{L} = 3000~\mathrm{fb}^{-1}$ and a representative combined
efficiency $\epsilon_{\rm sel} \approx 5\text{--}10\%$, estimated from
the geometric acceptance in the timing layer~\cite{CMS_MTD,ATLAS_HGTD},
the $\Delta t > 200~\mathrm{ps}$ signal fraction for
$M_{\rm LLP} \sim 300\text{--}600~\mathrm{GeV}$~\cite{Cheung:2021utb,Mason:2019okp},
and the displaced-lepton reconstruction efficiency from the MTD and HGTD
TDRs~\cite{CMS_MTD,ATLAS_HGTD}, the expected number of timing-tagged
$e/\mu$ displaced leptons---the $\tau$ channel, which carries no prompt
$e/\mu$ tag, is absorbed into $\epsilon_{\rm sel}$ as signal loss---is
$N_{\rm disp} = 2\,\sigma_{\rm NLO}\,\mathcal{L}\,\epsilon_{\rm sel}$,
where the factor of 2 counts the decay of \textit{both} $\tilde\eta^+$
and $\tilde\eta^-$ in each pair-production event:
$N_{\rm disp} \approx 1430\text{--}2856$, $424\text{--}846$, and
$64\text{--}128$ for the three mass points, respectively.

%-----------------------------------------------------------------------------
\subsection{Statistical discrimination}
\label{sec:statistics}
%-----------------------------------------------------------------------------

To distinguish NH from IH at $3\sigma$ confidence, we use an exact
one-sided Poisson test on the observed electron count $N_e$.
At $N_{\rm disp} = 25$, NH predicts $(N_e, N_\mu) = (3.1, 21.9)$ while IH
predicts $(17.0, 8.0)$; the exact Poisson probability of observing
$N_e \leq 3$ given the IH expectation of $17.0$ is
$p = 8.6 \times 10^{-5}$ ($3.8\sigma$), and the reverse
mis-classification probability is $p = 2.1 \times 10^{-7}$ ($5.1\sigma$).
Hence $\mathcal{O}(25)$ events provide $3\sigma$ discrimination in an
ideal Poisson counting experiment.
Systematic uncertainty on $r_\varepsilon$ at the few-percent level,
measurable from the displaced $J/\psi \to \ell^+\ell^-$ control sample of
Sec.~\ref{sec:robustness}, shifts $(N_e/N_\mu)^{\rm obs}$ by far less
than the factor $\Rdisc = 15.1$ separating the predictions and does not
materially raise this threshold.
A complementary binomial test, conditioning on the total $e+\mu$ count
and profiling over the $r_\varepsilon$ uncertainty, is presented in
Appendix~C; both statistical frameworks give
$\geq 3\sigma$ at $\mathcal{O}(10)$ events, confirming that the result is
robust to the choice of test.
Even the most challenging $600~\mathrm{GeV}$ benchmark accumulates
$N_{\rm disp} \approx 64$--$128$ and yields a Poisson $p$-value below
$0.003$ for the wrong hierarchy, confirming that the criterion of
Eq.~(\ref{eq:criterion}) is met across the full LLP mass range at HL-LHC
luminosity.

%-----------------------------------------------------------------------------
\subsection{Background considerations}
\label{sec:backgrounds}
%-----------------------------------------------------------------------------

A crucial feature of the flavor-\emph{ratio} observable is its resilience
to whatever prompt background survives the timing cut: because the
hierarchy is encoded in $N_e/N_\mu$ rather than in an absolute rate, a
flavor-symmetric residual---contributing equally to the $e$ and $\mu$
counts---can only drive the measured ratio toward unity and dilute the
discrimination, never reverse the direction of the NH--IH separation.
Only a flavor-\emph{asymmetric} residual can bias the assignment, and it
must therefore be controlled at the few-percent level using the same
displaced $J/\psi \to \ell^+\ell^-$ control sample that fixes
$r_\varepsilon$ (Sec.~\ref{sec:robustness}).
Under the conservative working assumption that the data-driven timing
sidebands suppress the prompt background to $\lesssim 10\%$ of the signal
with $e/\mu$ asymmetry below the few-percent level, the separation
degrades by less than $\mathcal{O}(1)$ and the decision criterion of
Eq.~(\ref{eq:criterion}) is preserved; a quantitative background estimate
with a full detector simulation is left to a dedicated experimental
study.

%=============================================================================
\section{Summary}
\label{sec:summary}
%=============================================================================

The key structural observation of this Letter is that the
$\delta_{CP}$--hierarchy degeneracy that afflicts every long-baseline
oscillation program is absent in the flavor-tomography observable:
$\Ne/\Nmu$ depends on $|U_{\alpha i}|^2$, not $U_{\alpha i}$, so no value
of $\delta_{CP}$ can shift the NH prediction ($0.140$) toward the IH
prediction ($2.11$) within the shared-coupling class.
The hierarchy is instead encoded in a ratio of lepton counts fixed
entirely by already-measured oscillation parameters, robust against
$\mathcal{O}(1)$ detector-efficiency variations, and insensitive to BSM
mass parameters.
The method is complementary to JUNO~\cite{JUNO:2024jaw,JUNO:2025gmd}:
both approaches are $\delta_{CP}$-independent and can operate in the same
era, but flavor tomography directly probes the Majorana coupling
structure of the LLP mediator, a dimension not addressed by oscillation
or reactor measurements.
We have verified the prediction across four representative BSM model
classes, analytically traceable to $2\tan^2\theta_{13}$ (NH) and
${\approx}1.9$ at maximal mixing (IH; exact value $2.11$), with a
discrimination ratio no smaller than $\Rdisc = 8.85$ across the full
$3\sigma$ oscillation band and $\Rdisc \geq 2.2$ within the
cosmologically allowed mass ranges~\cite{Planck2018,DESI:2024mwx}.
The prediction will only sharpen as oscillation measurements improve.

%=============================================================================
\begin{acknowledgments}
The author thanks colleagues at the Institute of High Energy Physics for
stimulating discussions.
\end{acknowledgments}

%=============================================================================
%\bibliographystyle{apsrev4-2}
\bibliography{references}

% ---- Appendices (placed after the references; titles hardcoded because
% ---- revtex4-2 disables \appendix numbering after \bibliography) ----

%=============================================================================
\section*{Appendix A: Dependence of $\Ne/\Nmu$ on the lightest neutrino mass}
%=============================================================================

\begin{figure}[t]
\centering
\includegraphics[width=0.9\columnwidth]{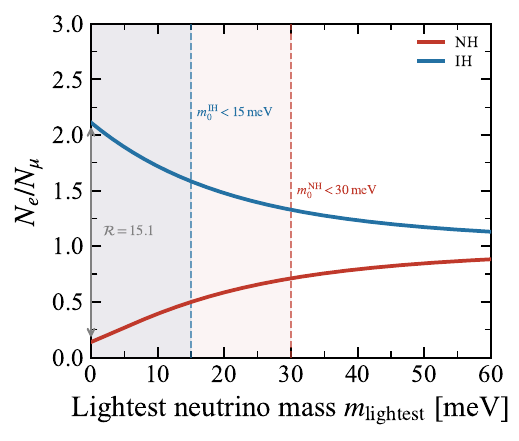}
\caption{Predicted $\Ne/\Nmu$ as a function of $m_{\rm lightest}$ for NH
(red) and IH (blue) at \NuFit\ best-fit mixing
angles~\cite{Esteban2020nufit}.
Vertical dashed lines: Planck 2018~\cite{Planck2018} upper bounds
(${<}30~\mathrm{meV}$ for NH, ${<}15~\mathrm{meV}$ for IH).
At the Planck boundaries the IH prediction is $1.58$ (at
$15~\mathrm{meV}$) and the NH prediction is $0.71$ (at
$30~\mathrm{meV}$), so the discrimination ratio satisfies
$\Rdisc \geq 2.2$ throughout the allowed region.
The critical mass $m_*(f{=}2) \approx 27~\mathrm{meV}$ lies within the
Planck NH-allowed range.
Both hierarchies approach the common value $1$ only far beyond the
Planck-allowed region; within it the IH curve decreases from $2.11$ at
$m_{\rm lightest}=0$ to ${\approx}1.58$ at its Planck edge, while the NH
curve rises from $0.140$ toward $0.71$.}
\label{fig:mlightest}
\end{figure}

Figure~\ref{fig:mlightest} shows $\Ne/\Nmu$ versus $m_{\rm lightest}$.
Both hierarchies converge toward the common asymptotic value
$\Ne/\Nmu \to 1$ as the spectrum becomes quasi-degenerate
($m_{\rm lightest} \gg 50~\mathrm{meV}$), because equal-mass eigenstates
weight every flavor by the unitary row sum
$\sum_i |U_{\alpha i}|^2 = 1$, so the flavor-weighted mass sums
$\mathcal{A}_e$ and $\mathcal{A}_\mu$ become equal.
We adopt the conservative Planck 2018 bound~\cite{Planck2018}
$\sum m_i < 120~\mathrm{meV}$ (the DESI BAO
constraint~\cite{DESI:2024mwx} is tighter but cosmology-combination
dependent).
Within this window the IH prediction stays at or above ${\approx}1.58$
(reached at its Planck-allowed maximum
$m_{\rm lightest} \approx 15~\mathrm{meV}$), while the NH prediction
rises to at most $0.71$ (at $m_{\rm lightest} \approx 30~\mathrm{meV}$),
so the discrimination ratio satisfies $\Rdisc \geq 2.2$ throughout the
Planck-allowed range.
For $m_{\rm lightest} \gtrsim 15~\mathrm{meV}$ the
$m_{\rm lightest}$-dependent geometric-midpoint threshold,
$\sqrt{(N_e/N_\mu)^{\rm NH} (N_e/N_\mu)^{\rm IH}}$ evaluated at the
relevant $m_{\rm lightest}$, replaces the benchmark value $0.543$ in
Eq.~(\ref{eq:criterion}).
Defining $m_*(f)$ as the lightest mass at which the IH-to-NH ratio,
evaluated at a common $m_{\rm lightest}$, equals $f$, one finds
$m_*(f{=}2) \approx 27~\mathrm{meV}$ (just beyond the IH Planck-allowed
range) and $m_*(f{=}10) \approx 2.4~\mathrm{meV}$ (comfortably above
current NH oscillation fits~\cite{Esteban2020nufit} and the KATRIN
direct limit~\cite{KATRIN2025}).

%=============================================================================
\section*{Appendix B: Numerical validation of the shared-coupling approximations}
%=============================================================================

\textit{Non-degenerate right-handed-neutrino masses.}
In the Scotogenic model, non-degenerate right-handed-neutrino masses
introduce a loop-function weight
$\mathcal{F}(M_{N_i}/m_\eta)$~\cite{Ma2006}, violating condition (ii);
a scan over spectra with splittings up to $50\%$ confirms that the
discrimination ratio satisfies $\Rdisc > 9$ in all cases.

\textit{Complex $R$-matrix.}
For complex $R = e^{i\omega}$, violating condition (iii),
$\sum_i|y_{i\alpha}|^2$ acquires corrections
$\propto \sinh^2(\mathrm{Im}\,\omega_{ij})$.
The MEG~II bound~\cite{MEGII2025}
$\mathrm{BR}(\mu\to e\gamma) < 1.5\times10^{-13}$ implies
$|\mathrm{Im}\,\omega_{12}|\lesssim 2.5$ ($M_N\sim 1~\mathrm{TeV}$);
a scan over $10^5$ matrices satisfying this bound gives deviations of
$\mathcal{O}(10\%)$ in $\Ne/\Nmu$, far below the factor
$\Rdisc = 15.1$ separating the hierarchy predictions.
In the less-constrained $e\tau/\mu\tau$ subspace the deviations reach
$\mathcal{O}(30\%)$, still well within the gap.

\textit{$\tau$ contamination.}
Three suppression factors keep $\tau$-induced $e/\mu$ contamination at
$\lesssim 0.5\%$: the leptonic $\tau$ sub-branching ratios
($\approx 18\%$ and $17\%$), the $\mathcal{O}(10^{-2})$ probability that
a secondary lepton from $\tau\to\ell\nu\bar\nu$
($c\tau_\tau\approx 87~\mu\mathrm{m}$) passes the
$\Delta t>200~\mathrm{ps}$ cut, and standard displaced-track veto
requirements.

%=============================================================================
\section*{Appendix C: Binomial statistical test}
%=============================================================================

Conditioning on $N_{e\mu}$ (total $e+\mu$ events), the electron count
$K_e$ is binomial:
\begin{equation}
  K_e \mid N_{e\mu} \sim \mathrm{Binomial}(N_{e\mu},\; p_e^H),
  \quad
  p_e^H = \frac{r_\varepsilon R_H}{1+r_\varepsilon R_H},
  \tag{C1}
\end{equation}
with $R_{\rm NH}=0.140$, $R_{\rm IH}=2.11$, giving $p_e^{\rm NH}=0.123$
and $p_e^{\rm IH}=0.679$ at $r_\varepsilon=1$.
This conditioning on the total $e+\mu$ count is insensitive to the
overall signal yield, reducing the test to a pure flavor-fraction
comparison.
Table~\ref{tab:SM_significance} gives one-sided mis-classification
$p$-values; the minimum $N_{e\mu}$ for $3\sigma$ in \textit{both}
directions is $N_{e\mu}=9$.
Profiling over $r_\varepsilon$ as a nuisance parameter, even a
$\pm30\%$ uncertainty on $r_\varepsilon$ leaves $>4.9\sigma$ at
$N_{e\mu}=25$, because the factor-of-$5.5$ gap between $p_e^{\rm NH}$
and $p_e^{\rm IH}$ far exceeds any $\mathcal{O}(1)$ efficiency
variation.
Both this binomial test and the Poisson test of
Sec.~\ref{sec:statistics} give $\geq3\sigma$ at $\mathcal{O}(10)$
events, confirming that the result is robust to the choice of
statistical framework.

\begin{table}[t]
\centering
\caption{Binomial mis-classification $p$-values at $r_\varepsilon=1$,
\NuFit\ best-fit, $m_{\rm lightest}=0$ (left); worst-case significance,
in units of $\sigma$, after profiling over relative $r_\varepsilon$
uncertainties of $\pm 10\%$, $\pm 20\%$, and $\pm 30\%$ (right).}
\label{tab:SM_significance}
\footnotesize
\setlength{\tabcolsep}{2pt}
\begin{tabular}{ccccccc|ccc}
\hline
$N_{e\mu}$ & $K_e^{\rm NH}$ & $K_e^{\rm IH}$ &
  $p_{\rm NH\to IH}$ & $\sigma$ &
  $p_{\rm IH\to NH}$ & $\sigma$ &
  $10\%$ & $20\%$ & $30\%$ \\
\hline
10 & 1 & 7  & $7.3\!\times\!10^{-4}$ & 3.2 & $2.3\!\times\!10^{-4}$ & 3.5
   & 3.2 & 3.1 & 3.0 \\
15 & 2 & 10 & $6.5\!\times\!10^{-6}$ & 4.1 & $3.4\!\times\!10^{-6}$ & 4.7
   & 4.1 & 4.0 & 3.9 \\
25 & 3 & 17 & $1.1\!\times\!10^{-8}$ & 5.6 & $1.8\!\times\!10^{-10}$ & 6.3
   & 5.3 & 5.1 & 4.9 \\
\hline
\end{tabular}
\end{table}

%=============================================================================
\end{document}